\documentclass[twocolumn,showpacs,superscriptaddress,prl]{revtex4}

\usepackage{amsmath,amssymb} 
\usepackage{graphicx}

\begin{document}

\title{Quasi-relativistic behavior of cold atoms in light fields}

\date{\today{}}

\author{Gediminas Juzeli\={u}nas}
\affiliation{Institute of Theoretical Physics and Astronomy of Vilnius
University, A. Go\v{s}tauto 12, Vilnius 01108, Lithuania}

\author{Julius Ruseckas}
\affiliation{Institute of Theoretical Physics and Astronomy of Vilnius
University, A. Go\v{s}tauto 12, Vilnius 01108, Lithuania}

\author{Markus Lindberg}
\affiliation{Department of Physics, \AA bo Akademi University, \AA bo FIN-20500,
Finland}

\author{Luis Santos}
\affiliation{Institut f\"ur Theoretische Physik, Leibniz Universit\"at, Hannover
D30167, Germany}

\author{Patrik \"Ohberg}
\affiliation{SUPA, School of Engineering and Physical Sciences, Heriot-Watt
University, Edinburgh EH14 4AS, United Kingdom}

\begin{abstract}
We study the influence of three laser beams on the center of mass motion of
cold atoms with internal energy levels in a tripod configuration. We show that
similar to electrons in graphene the atomic motion can be equivalent to the
dynamics of ultra-relativistic two-component Dirac fermions. We propose and
analyze an experimental setup for observing such a quasi-relativistic motion of
ultracold atoms. We demonstrate that  the atoms can experience negative
refraction and focussing by Veselago-type lenses. We also show how the chiral
nature of the atomic motion manifests itself as an oscillation of the atomic
internal state population which depends strongly on the direction of the center
of mass motion. For certain directions an atom remains in its initial state,
whereas for other directions the populations undergo oscillations between a
pair of internal states.
\end{abstract}


\maketitle

\section{Introduction}

Two-dimensional (2D) quantum systems are a source of many remarkable phenomena.
A striking example in this respect is provided by the properties of electrons
in graphene
\cite{novoselov05,mccann06,novoselov06,geim07,novoselov07,matulis07,jackiw07,pendry07,cheianov07}
--- a two dimensional hexagonal crystal of carbon atoms.  Near the Fermi level
the electrons in graphene behave like massless ultra-relativistic two-component
Dirac fermions \cite{beresteckii82} moving with a velocity $v_F$ which does not
depend on the momentum. This leads to a number of distinct effects in graphene
and graphene bilayers, such as a half integer quantum Hall effect
\cite{novoselov05,geim07,novoselov07} and the Klein paradox \cite{novoselov06}.
Furthermore it has been suggested recently that electrons in graphene should
exhibit a negative refraction \cite{pendry07,cheianov07} at a potential
barrier, similar to the electromagnetic waves refracting in a counterintuitive
way at an interface with a material characterized by a negative dielectric
permitivity and a negative magnetic permeability \cite{veselago68,pendry00}.

In this Rapid Communication we show how cold atoms obtain ultra-relativistic properties of
Dirac fermions if manipulated by laser beams. We suggest to use three laser
beams acting on atoms in a tripod configuration \cite{unanyan98,ruseckas05}.
The light beams induce an effective vector potential (the Mead-Berry connection
\cite{berry84,wilczek84,mead91}) which influences the atomic center of mass
motion \cite{ruseckas05}. We demonstrate that by choosing proper light fields
the vector potential can be made proportional to an operator of spin $1/2$. For
small momenta the atomic motion becomes equivalent to the ultra-relativistic
motion of two-component Dirac fermions, as is the case for electrons in
graphene. We propose and analyze an experimental setup for observing such a
quasi-relativistic behavior of the cold atoms. We show that the atoms can
experience negative refraction and focussing by Veselago-type lenses
\cite{veselago68}. 

Interestingly the velocity of the quasi-relativistic atoms is of the order of a
centimeter per second. This is ten orders of magnitude smaller than the speed
of light in vacuum $c\approx3\times10^8\,\mathrm{m/s}$.  For comparison, the
velocity of the Dirac-type electrons in graphene
$v_F\approx 10^6\,\mathrm{m/s}$ is only three hundred times smaller
than $c$ \cite{geim07}. Thus the ultra-relativistic behavior of cold atoms
manifests itself at extremely small velocities.  Note also that our proposal
does not need a lattice and thus operates in a continuous regime in contrast to
recent papers on the quasi-relativistic dynamics of cold atoms in
one-dimensional \cite{ruostekoski02} or 2D hexagonal (graphene-type)
\cite{zhu07} lattices.

\section{Formulation}

\begin{figure}
\centering
\includegraphics[width=0.45\textwidth]{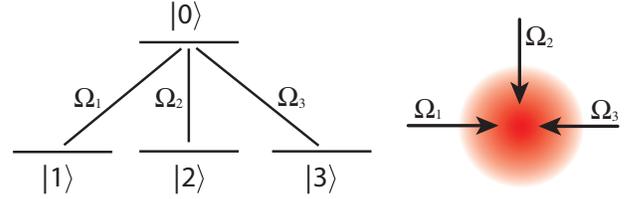}
\caption{(Color online) An atom interacting with three laser fields in a tripod
configuration.}
\label{fig1}
\end{figure}

Let us consider the adiabatic motion of atoms in the field of three stationary
laser beams. The beams couple four internal atomic levels in a tripod
configuration \cite{unanyan98,ruseckas05}, in which\ the atoms are
characterized by three lower levels $|1\rangle$, $|2\rangle$ and $|3\rangle$
and an excited level $|0\rangle$ shown in Fig.~\ref{fig1}a
\cite{footnote-transitions}.  The $j$-th laser induces a transition (with a
Rabi frequency $\Omega_j$) between the $j$-th lowest level and the excited
level $|0\rangle$.  We shall be interested in a scheme where the first two
lasers have the same intensities and counter-propagate in the $x$-direction
while the third one propagates in the negative $y$-direction (see
Fig.~\ref{fig1}b), i.e.\ $\Omega_1=\Omega\sin\theta\mathrm{e}^{-i\kappa
x}/\sqrt{2}$, $\Omega_2=\Omega\sin\theta\mathrm{e}^{i\kappa x}/\sqrt{2}$ and
$\Omega_3=\Omega\cos\theta\mathrm{e}^{-i\kappa y}$, where
$\Omega=\sqrt{|\Omega_1|^2+|\Omega_2|^2+|\Omega_3|^2}$ is the total Rabi
frequency, and the mixing angle  $\theta$ defines the relative intensity.

The electronic Hamiltonian of the tripod atom reads in the interaction
representation \cite{ruseckas05}
\begin{equation}
\hat{H}_0=-\hbar\Bigl(\Omega_1|0\rangle\langle1|+\Omega_2|0\rangle\langle2|
+\Omega_3|0\rangle\langle3|\Bigr)+\mathrm{H.c.}\,.
\label{eq:H-0}
\end{equation}
The Hamiltonian $\hat{H}_0$ has two eigen-states of zero energy containing no
contribution from the excited state $|0\rangle$: 
\begin{eqnarray}
|D_1\rangle & = &\frac{1}{\sqrt{2}}e^{-i\kappa y}\left(e^{i\kappa
x}|1\rangle-e^{-i\kappa x}|2\rangle\right)
\label{eq:D1-plane-wave}
\\
|D_2\rangle & = &\frac{1}{\sqrt{2}}e^{-i\kappa y}\cos\theta\left(e^{i\kappa
x}|1\rangle+e^{-i\kappa
x}|2\rangle\right)-\sin\theta|3\rangle
\label{eq:D2-plane-wave}
\end{eqnarray}
The states  $|D_1\rangle$ and  $|D_2\rangle$  known as the dark states
\cite{unanyan98,ruseckas05} depend on the atomic position through the spatial
dependence of the Rabi-frequencies $\Omega_j$. 

The adiabatic approximation is carried out neglecting transitions from the dark
states to the bright state
$|B\rangle\sim\Omega_1^*|1\rangle+\Omega_2^*|2\rangle+\Omega_3^*|3\rangle$. The
latter is coupled to the excited atomic state $|0\rangle$ with the Rabi
frequency $\Omega$, so the two states $|B\rangle$ and $|0\rangle$ split into a
doublet separated from the dark states by the energies $\pm\Omega$. The
adiabatic approximation is justified if $\Omega$ is sufficiently large compared
to the two-photon detuning due to the laser mismatch and/or Doppler shift. In
that case the internal state of an atom evolves within the dark state manifold.
The atomic state-vector $|\Phi\rangle$ can then be expanded in terms of the
dark states according to
$|\Phi\rangle=\sum_{j=1}^2\Psi_j(\mathbf{r})|D_j(\mathbf{r})\rangle$, where
$\Psi_j(\mathbf{r})$ is the wave-function for the center of mass motion of the
atom in the $j$-th dark state. 

Thus atomic center of mass motion is described by a two-component wave-function
$\Psi=(\Psi_1,\Psi_2)^{T}$.  The column-matrix $\Psi$ obeys the
Schr\"odinger equation \cite{ruseckas05} 
\begin{equation}
i\hbar\frac{\partial}{\partial
t}\Psi=\left[\frac{1}{2m}(-i\hbar\nabla-\mathbf{A})^2+V+\Phi\right]\Psi,
\label{eq:SE-reduced}
\end{equation}
where $m$ is the atomic mass, and $\mathbf{A}$, $\Phi$ and $V$ are $2\times2$
matrices. The gauge potentials $\mathbf{A}$ and $\Phi$ emerge due to the
spatial dependence of the atomic dark states. The reduced $2\times2$ matrix
$\mathbf{A}$ with the elements $\mathbf{A}_{n,m}=i\hbar\langle
D_n(\mathbf{r})|\nabla D_m(\mathbf{r})\rangle$ represents the effective vector
potential known as the Mead-Berry connection
\cite{ruseckas05,berry84,wilczek84,mead91}.  The $2\times2$ matrix $\Phi$ acts
as an effective scalar potential. The external potential $V$ confines the
motion of the dark-state atoms to a finite region in space. Specifically we
have $V_{n,m}=\langle D_n(\mathbf{r})|\hat{V}|D_m(\mathbf{r})\rangle$ with
$\hat{V}=V_1(\mathbf{r})|1\rangle\langle1|+V_2(\mathbf{r})|2\rangle\langle2|
+V_2(\mathbf{r})|3\rangle\langle3|$, where $V_j(\mathbf{r})$ is the trapping
potential for an atom in the internal state $j=1,2,3$. Note that the potential
$V_j$ can also accommodate a possible detuning of the $j$-th laser from the
resonance of the $j\rightarrow0$ transition.

\section{Atomic motion in the field of three plane waves }

The potentials $\mathbf{A}$, $\Phi$ and $V$ have been considered in
Ref.~\cite{ruseckas05} for arbitrary light fields. In the present configuration
of the light fields, the potentials take the form 
\begin{eqnarray}
\mathbf{A} & = &\hbar\kappa\left(
\begin{array}{cc}
\mathbf{e}_y & -\mathbf{e}_x\cos\theta\\ -\mathbf{e}_x\cos\theta &
\mathbf{e}_y\cos^2\theta
\end{array}\right)\,,
\label{eq:A-Matrix}
\\
\Phi & = &\left(
\begin{array}{cc}
\hbar^2\kappa^2\sin^2\theta/2m & 0\\ 0 &
\hbar^2\kappa^2\sin^2(2\theta)/8m
\end{array}\right)\,,
\label{eq:Pfi-Matrix}
\\
V & = &\left(
\begin{array}{cc}
V_1 & 0\\ 0 & V_1\cos^2\theta+V_3\sin^2\theta\end{array}\right)\,,
\label{eq:V-Matrix}
\end{eqnarray}
where the external trapping potential is assumed to be the same for the first
two atomic states, $V_1=V_2$.

In what follows we take $V_3-V_1=\hbar^2\kappa^2\sin^2(\theta)/2m$. This can be
achieved by detuning the third laser from the two-photon resonance by the
frequency $\Delta\omega_3=\hbar\kappa^2\sin^2\theta/2m$. Thus the overall
trapping potential simplifies to $V+\Phi=V_1\mathrm{I}$ (up to a constant),
where $\mathrm{I}$ is the unit matrix. In other words, both dark states are affected by
the same trapping potential $V_1\equiv V_1(\mathbf{r})$.

Furthermore we take the mixing angle $\theta=\theta_0$ to be such that
$\sin^2\theta_0=2\cos\theta_0$, giving $\cos\theta_0=\sqrt{2}-1$. In that case
the vector potential can be represented in a symmetric way in terms of the
Pauli matrices $\sigma_x$ and $\sigma_z$
\begin{equation}
\mathbf{A}=\hbar\kappa^{\prime}(-\mathbf{e}_x\sigma_x+\mathbf{e}_y\sigma_z)
+\hbar\kappa_0\mathbf{e}_y\mathrm{I},
\label{eq:A-Compact-Symmetric}
\end{equation}
where $\kappa^{\prime}=\kappa\cos\theta_0\approx0.414\kappa$ and
$\kappa_0=\kappa(1-\cos\theta_0)$. Although the vector potential is constant,
it can not be eliminated via a gauge transformation, because the Cartesian
components $A_x$ and $A_y$ do not commute. Thus the light-induced potential
$\mathbf{A}$ is non-Abelian.  Note that non-Abelian gauge potentials can also
be induced in optical lattices using other techniques \cite{osterloh05}. 

It is convenient to introduce new dark states: 
\begin{eqnarray}
|D_1^{\prime}\rangle & = &
\frac{1}{\sqrt{2}}\left(|D_1\rangle+i|D_2\rangle\right)e^{i\kappa_0y},
\label{eq:D1-transformed}
\\ |D_2^{\prime}\rangle & = &
\frac{i}{\sqrt{2}}\left(|D_1\rangle-i|D_2\rangle\right)e^{i\kappa_0y}.
\label{eq:D2-transformed}
\end{eqnarray}
The transformed two component wave-function is related to the original one
according to
$\Psi^{\prime}=\exp(-i\kappa_0y)\exp\left(-i\frac{\pi}{4}\sigma_x\right)\Psi$.
The exponential factor $\exp(-i\kappa_0y)$ induces a shift in the origin of the
momentum $\mathbf{k}\rightarrow\mathbf{k}-\kappa_0\mathbf{e}_y$. With the new
set of dark states we get the vector potential
$\mathbf{A}^{\prime}=-\hbar\kappa^{\prime}\mathbf{\sigma}_{\bot}$, where
$\mathbf{\sigma}_{\bot}=\mathbf{e}_x\sigma_x+\mathbf{ e}_y\sigma_y$ is the spin
$1/2$ operator in the $xy$ plane.  The transformed equation of the atomic
motion takes the form 
\begin{equation}
i\hbar\frac{\partial}{\partial
t}\Psi^{\prime}=\left[\frac{1}{2m}(-i\hbar\nabla+\hbar\kappa^{\prime}\mathbf{
\sigma}_{\bot})^2+V_1\right]\Psi^{\prime}.
\label{eq:SE-Transformed-prime}
\end{equation}
In this way the vector potential governing the atomic motion is proportional to
the spin operator $\mathbf{\sigma}_{\bot}$.

If the trapping potential $V_1$ is constant, we can consider plane-wave
solutions,
\begin{equation}
\Psi^{\prime}(\mathbf{r},t)=\Psi_{\mathbf{k}}e^{i\mathbf{k}\cdot\mathbf{r}
-i\omega_{\mathbf{k}}t},\qquad\Psi_{\mathbf{k}}=\left(
\begin{array}{c}
\Psi_{1\mathbf{k}}\\\Psi_{2\mathbf{k}}\end{array}\right),
\label{eq:Fourier-Transform}
\end{equation}
where $\omega_{\mathbf{k}}$ is the eigen-frequency. The $\mathbf{k}$-dependent
spinor $\Psi_{\mathbf{k}}$ obeys the stationary Schr\"odinger equation
$H_{\mathbf{k}}\Psi_{\mathbf{k}}=\hbar\omega_{\mathbf{k}}\Psi_{\mathbf{k}},$
with the following $\mathbf{k}$-dependent Hamiltonian
\begin{equation}
H_{\mathbf{k}}=\frac{\hbar^2}{2m}(\mathbf{k}+\kappa^{\prime}\mathbf{\mathbf{
\sigma}_{\bot}})^2+V_1.
\label{eq:H-k}
\end{equation}
For small wave-vectors ($k\ll\kappa^{\prime}$) the atomic Hamiltonian reduces
to the Hamiltonian for the 2D relativistic motion of a two-component massless
particle of the Dirac type known also as the Weyl equation \cite{maggiore05},
\begin{equation}
H_{\mathbf{k}}=\hbar
v_0\mathbf{k}\cdot\mathbf{\sigma}_{\bot}+V_1+mv_0^2,
\label{eq:H-k-1a}
\end{equation}
where $v_0=\hbar\kappa^{\prime}/m$ is the velocity of such a quasi-relativistic
particle. The velocity $v_0$ represents the recoil velocity corresponding to
the wave-vector $\kappa^{\prime}$, and is typically of the order of a
centimeter per second.

The Hamiltonian $H_{\mathbf{k}}$ commutes with the $2D$ chirality operator
$\sigma_{\mathbf{k}}=\mathbf{k}\cdot\mathbf{\mathbf{\sigma}_{\bot}}/k$.  The
later has the eigenstates
\begin{equation}
\Psi_{\mathbf{k}}^{\pm}=\frac{1}{\sqrt{2}}\left(
\begin{array}{c}
1\\\pm\frac{k_x+ik_y}{k}\end{array}\right),
\label{eq:Psi-k-pm}
\end{equation}
for which
$\sigma_{\mathbf{k}}\Psi_{\mathbf{k}}^{\pm}=\pm\Psi_{\mathbf{k}}^{\pm}$.  Here
the chirality is associated with the subspace of the dark states defined by the
basis in Eqs.~(\ref{eq:D1-transformed})--(\ref{eq:D2-transformed}) rather than
with the spin states in the usual sense. The corresponding dynamics can
therefore also be illustrated by for instance a Poincar\'e sphere.

\begin{figure}
\centering
\includegraphics[width=0.45\textwidth]{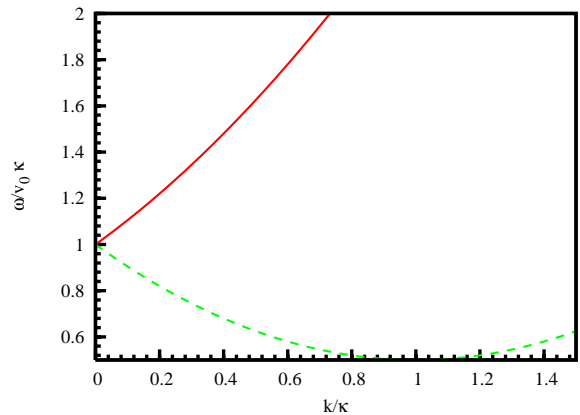}
\caption{(Color online) Upper (red solid) and lower (green dashed) dispersion
branches for a tripod atom in light fields}
\label{fig2}
\end{figure}

The eigenstates (\ref{eq:Psi-k-pm}) are also eigenstates of the Hamiltonian
$H_{\mathbf{k}}$ with eigen-frequencies
$\omega_{\mathbf{k}}\equiv\omega_{\mathbf{k}}^{\pm}$. In what follows the
atomic motion is assumed to be confined in the $xy$ plane. The dispersion then
reads
\begin{equation}
\hbar\omega_{\mathbf{k}}^{\pm}=\hbar
v_0(k^2/2\kappa^{\prime}\pm
k)+V_1+mv_0^2\,,
\label{eq:eigenvalue-1-specific-1}
\end{equation}
see Fig.~\ref{fig2} in which $V_1=-mv_0^2/2$. The atomic motion in different
dispersion branches is characterized by opposite chirality if the direction
$\mathbf{k}/k$ is fixed.  For small wave-vectors ($k\ll\kappa^{\prime}$) the
dispersion simplifies to $\hbar\omega_{\mathbf{k}}^{\pm}=\pm\hbar v_0k+V_1+mv_0^2$, where
the upper (lower) sign corresponds to a linear cone with a positive (negative)
group velocity, $v_g^{\pm}=\pm v_0$. Exactly the same dispersion is featured
for electrons near the Fermi level in graphene
\cite{novoselov05,mccann06,novoselov06,geim07,novoselov07}.

\section{Proposed experiment}

To observe the quasi-relativistic behavior of cold atoms, we propose the
following experimental situation. We suppose initially an atom (or a dilute
atomic cloud) is in the internal state $|3\rangle$ with a translational motion
described by a wave-packet with a central wave-vector
$\mathbf{k}_{\mathrm{in}}$ and a wave-vector spread $\Delta k\ll
k_{\mathrm{in}}$, i.e.\ $|\Psi_{\mathrm{in}}
\rangle=\psi(\mathbf{r})e^{i\mathbf{k}_{\mathrm{in}}\cdot\mathbf{r}}
|3\rangle$, where the envelope function $\psi(\mathbf{r})$ varies slowly within
the wavelength $\lambda_{\mathrm{in}}=2\pi/k_{\mathrm{in}}$.  The cold atoms
can be set in motion using various techniques, e.g.\ by means of the two-photon
scattering which induces a recoil momentum
$\hbar\mathbf{k}_{in}=\hbar\mathbf{k}_{\mathrm{2phot}}$ to the atoms, where
$\mathbf{k}_{\mathrm{2phot}}$ is a wave-vector of the two-photon mismatch
\cite{Deng99}.

Initially all three lasers are off. Subsequently the lasers are switched on in
a counterintuitive manner, switching the lasers $1$ and $2$ on first followed
by the laser $3$. At the beginning of this stage the internal state $|3\rangle$
coincides with the dark state $|D_2\rangle$, so the original and transformed
multi-component wave-functions are given by:
\begin{equation}
\Psi=\left(
\begin{array}{c}
0\\ 1\end{array}\right)\psi(\mathbf{r})e^{i\mathbf{k}_{\mathrm{in}}
\cdot\mathbf{r}},
\quad
\Psi^{\prime}=\frac{1}{\sqrt{2}}\left(
\begin{array}{c}
-i\\ 1\end{array}\right)\psi(\mathbf{r})e^{i\mathbf{k}\cdot\mathbf{r}},
\label{eq:psi-D-prime-setup}
\end{equation}
where the transformed wave-function $\Psi^{\prime}$ is obtained expressing the
original dark state $|D_2\rangle$ through the new ones
$|D_{1,2}^{\prime}\rangle$, with
$\mathbf{k}=\mathbf{k}_{\mathrm{in}}-\kappa_0\mathbf{e}_y$ being a new central
wave-vector. If the laser $3$ is switched on sufficiently slowly, the atom
remains in the dark state $|D_2\rangle$ during the whole switch-on stage. Yet
the duration of the switching on should be short enough to prevent the dynamics
of the atomic center of mass at this stage.  To have ultra-relativistic
behaving atoms, the wave-number $k$ should be small $k\ll\kappa$, so that
$\mathbf{k}$ is a small contribution to
$\mathbf{k}_{\mathrm{in}}=\kappa_0\mathbf{e}_y+\mathbf{k}$. In addition, the
wave-number spread $\Delta k \ll k$, i.e.\ the width of the atomic wave-packet
is much larger than the central wave-length.  The subsequent centre of mass
motion of atoms in the laser fields is sensitive to the direction of
the wave-vector $\mathbf{k}$.

(i) If $\mathbf{k}=\pm k\mathbf{e}_y$, the wave-function
(\ref{eq:psi-D-prime-setup}) reads:
\begin{equation}
\Psi^{\prime}=-i\Psi_{\mathbf{k}}^{\pm}\psi(\mathbf{r})e^{\pm
iky},\qquad\Psi_{\mathbf{k}}^{\pm}=\frac{1}{\sqrt{2}}\left(
\begin{array}{c}
1\\ i\end{array}\right).
\label{eq:psi-D-prime-k-0--y}
\end{equation}
The upper (lower) sign in $\mathbf{k}=\pm k\mathbf{e}_y$ corresponds to a
situation where the atom is characterized by a positive (negative) chirality,
hence being in the upper (lower) dispersion branch. In both cases the atomic
wave-packet propagates along the $y$ axis with the velocity
$\mathbf{v}_0=\mathbf{e}_y\hbar\kappa^{\prime}/m$.

(ii) If the wave-vector is along the $x$ axis ($\mathbf{k}=k\mathbf{e}_x$), the
multi-component wave-function (\ref{eq:psi-D-prime-setup}) takes the form
\begin{equation}
\Psi^{\prime}=(c_{+}\Psi_{\mathbf{k}}^{+}+c_{-}\Psi_{\mathbf{k}}^{-})
\psi(\mathbf{r})e^{i\mathbf{k}\cdot\mathbf{r}},
\quad\Psi_{\mathbf{k}}^{\pm}=\frac{1}{\sqrt{2}}\left(
\begin{array}{c}
1\\\pm1\end{array}\right),
\label{eq:psi-D-prime-k-0--x}
\end{equation}
where $c_{\pm}=(-i\pm 1)/2$. In that case the initial wave-packet splits into
two with equal weights ($|c_{\pm}^2|=1/2$) and the same wave-vector
$\mathbf{k}$.  The wave-packets are characterized by different chiralities
and thus move in opposite directions with the velocities 
$\mathbf{v}_0=\pm\mathbf{e}_x\hbar\kappa^{\prime}/m$. 

Suppose the time is sufficiently small ($v_0t<d$), so the wave-packets of the
width $d$ are not yet spatially separated. The internal atomic state will then
undergo temporal oscillations between the dark states $|D_2\rangle$ and
$|D_1\rangle$ with a frequency equal to
$\omega_{\mathbf{k}}^{+}-\omega_{\mathbf{k}}^{-}=2v_0k$. Such an internal
dynamics can be detected by switching the laser $3$ off at a certain time. This
transforms the dark state $|D_2\rangle$ to the physical state $|3\rangle$.
Subsequently one can measure population of the state $|3\rangle$ for various
delay times and various wave-vectors $\mathbf{k}$. The chiral nature of the
atomic motion will manifest itself in the oscillations of the population of the
atomic state $|3\rangle$ if $\mathbf{k}$ is along the $x$ axis, and the absence
of such oscillations if $\mathbf{k}$ is along the $y$ axis.

\begin{figure}
\centering
\includegraphics[width=0.3\textwidth]{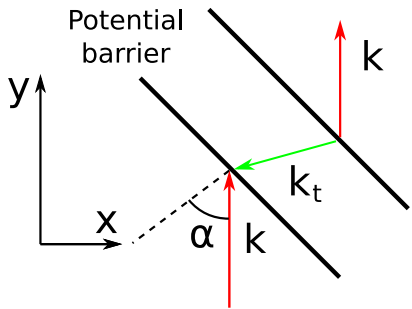}
\caption{(Color online) Negative refraction of cold atoms at a potential
barrier.  The incoming and outgoing atoms are in the upper (red) dispersion
branch, whereas the atoms inside the barrier are in the lower (green) one.}
\label{fig3}
\end{figure}

Furthermore as a consequence of the constructed Hamiltonian (\ref{eq:H-k-1a}),
the quasi-relativistic atoms can show negative refraction at a potential
barrier and thus exhibit focussing by Veselago-type lenses
\cite{veselago68,pendry00}.  Consider incident atoms that are in the upper
dispersion branch and propagate along the $y$ axis with a wave-vector
$\mathbf{k}=k\mathbf{e}_y$. Let us place a potential barrier of a height
$2\hbar v_0k$ at an angle of incidence $\alpha$  (see Fig.~\ref{fig3}). Inside
the barrier the atoms are transferred to the lower dispersion branch with
$\mathbf{k}_t=-k[\cos(2\alpha)\mathbf{e}_y+\sin(2\alpha)\mathbf{e}_x]$. This
would lead to the negative refraction of cold atoms at the barrier as shown in
Fig.~\ref{fig3}. Thus the potential barrier can act as a flat lens which
refocuses the atomic wave-packet.

In summary we have shown how the atomic motion can be equivalent to the
dynamics of ultra-relativistic (massless) two-component Dirac fermions.  As a
result the ultracold atoms can experience negative refraction and focusing by
Veselago-type lenses. In addition, we have investigated another manifestation of
the chiral nature of the atomic motion through dynamics of the population of
the internal atomic states.

\begin{acknowledgments}
This work was supported by  DFG (SFB407, SPP1116), the Royal Society of Edinburgh
and the UK Engineering and Physical Sciences Research Council. We thank
A.~Matulis for stimulating discussions on graphene, and D.~Boiron, J.~Arlt and
E.~Tiemann for information about experimental realizations.
\end{acknowledgments}

\end{document}